\documentclass[aps,prl,twocolumn,showpacs,superscriptaddress,groupedaddress,nofootinbib]{revtex4}  
\usepackage{graphicx}  
\usepackage{dcolumn}   
\usepackage{bm}        
\usepackage{bbold}
\usepackage{amssymb}   
\usepackage{amsmath}
\usepackage{epsfig}
\usepackage{multirow}
\usepackage{color}
\usepackage[normalem]{ulem}

\newcommand{\gsim}{\raise.3ex\hbox{$>$\kern-.75em\lower1ex\hbox{$\sim$}}}
\newcommand{\lsim}{\raise.3ex\hbox{$<$\kern-.75em\lower1ex\hbox{$\sim$}}}

\newcommand{\SO}{\text{SO}}
\newcommand{\SU}{\text{SU}}
\newcommand{\U}{\text{U}}
\newcommand{\Sp}{\text{Sp}}

\begin{document}
\title{Singlets in Composite Higgs Models in light of the LHC di-photon searches}

\author{Alexander Belyaev}
\affiliation{ School of Physics \& Astronomy, University of Southampton, UK}
\affiliation{Particle Physics Department, Rutherford Appleton Laboratory, 
Chilton, Didcot, Oxon OX11 0QX, UK}
\author{Giacomo~Cacciapaglia}
\author{Haiying~Cai}
\affiliation{Universit\'{e} de Lyon, France; Universit\'{e} Lyon 1, Villeurbanne, France}
\affiliation{CNRS/IN2P3, UMR5822, IPNL F-69622 Villeurbanne Cedex, France}
\author{Thomas~Flacke}
\affiliation{Department of Physics, Korea University, Seoul 136-713, Korea.}
\author{Alberto~Parolini}
\affiliation{Quantum Universe Center, Korea Institute for Advanced Study, Seoul 130-722, Korea.}
\author{Hugo Ser\^{o}dio}
\affiliation{Department of Physics, Korea University, Seoul 136-713, Korea.}


\date{\today}

\begin{abstract}
Models of compositeness can successfully address the origin of the Higgs boson, as a pseudo Nambu Goldstone boson (pNGB) of
a spontaneously broken global symmetry, and flavour physics via the partial compositeness mechanism. If the dynamics is
generated by a confining gauge group with fermionic matter content, there exists only a finite set of models that have the
correct properties to account for the Higgs and top partners at the same time. In this letter we explore the theory
space of this class of models: remarkably, all of them  contain -- beyond the pNGB Higgs -- a pNGB singlet, $ a $,
which couples to Standard Model gauge bosons via Wess-Zumino-Witten interactions, thus providing naturally a resonance in
di-boson at the LHC. With the assumption that the recently reported di-photon excess at 750 GeV at the LHC arises from the
$ a $-resonance,   we propose a generic approach on how to delineate the best candidate for composite Higgs models with
top-partners. We find that constraints from other di-boson searches severely reduce the theory space of the models under
consideration. For the models which can explain the di-photon excess, we make precise and testable predictions for the width
and other di-boson resonance searches.

\end{abstract}


\maketitle


\section{Introduction}
The idea that the breaking of the electroweak symmetry is due to a confining gauge theory is as old as the Standard Model itself~\cite{Weinberg:1975gm,Susskind:1978ms}.
The composite nature of the Higgs boson, therefore, defines a well motivated framework to address the Standard Model (SM) hierarchy problem. In modern composite Higgs models the Higgs doublet is realised as a pseudo-Nambu-Goldstone boson (pNGB)~\cite{Kaplan:1983fs,Kaplan:1983sm} of an approximate global symmetry broken by a new physics sector at a scale $f\sim1$ TeV.  
Inspired by constructions in warped extra dimensions~\cite{Agashe:2004rs}, the mass for the top is generated via partial compositeness~\cite{Kaplan:1991dc}, i.e. a linear mixing of the elementary fermions with composite top partners. 
In this letter, we  focus on four dimensional theories of this type, arising from an underlying confining gauge theory of interacting fermions, barring the presence of elementary scalars. The new Hyper-Colour (HC) gauge group $G_{HC}$ becomes strongly coupled in the infrared, and a phase transition is expected at a scale $\Lambda_{HC} \sim 4 \pi f$, around $10$ TeV, where a set of resonances should appear as composite states.

Building an underlying theory that contains both a composite Higgs and composite top partners is not an easy task, as many conditions need to be satisfied. In~\cite{Barnard:2013zea,Ferretti:2013kya,Ferretti:2014qta}, a first attempt in this direction was made: a key observation is that only a finite list of models in terms of HC groups and fermion representations is possible. Furthermore, the requirement of coupling the HC dynamics to ordinary gluons prefers models where there are two species of fermions transforming under two different representations of the HC group (for an example with a single species, see~\cite{Vecchi:2015fma}).
We would like to point out in this letter that a generic prediction of this class of models is the presence of additional pNGBs, besides the Higgs, which are parametrically lighter than other composite states.
The presence of more than one species of underlying fermions also allows to define global U(1)'s which are spontaneously broken, thus resulting in pNGBs that are singlets under the SM gauge group.

In models with only two species of fundamental fermions, there is a single non-anomalous $\U(1)$, independently on the representation of the fermions under $G_{HC}$. As the SM couplings do not break such $\U(1)$ (for the fermions, it suffices that all the top partners have the same charge), the mass of the singlet can have any value as it is unrelated to the mechanism of electroweak symmetry breaking. Furthermore it couples singly to a pair of gauge bosons via the Wess-Zumino-Witten (WZW)~\cite{Wess:1971yu,Witten:1979vv} anomaly.
This property is tantalising in view of recent observations at the LHC of excesses in di-boson searches, that may be explained as a scalar resonance~\cite{Cacciapaglia:2015nga}. 
Among such hints there were recent reports from ATLAS~\cite{ATLAS} and CMS~\cite{CMS} collaborations on the excess in the di-photon invariant mass around 750 GeV with local significance 3.9 and 2.6 standard deviations respectively. The models we consider here are qualitatively different from other proposals in the literature~\cite{Harigaya:2015ezk,Nakai:2015ptz,Franceschini:2015kwy,Bellazzini:2015nxw,Molinaro:2015cwg,Matsuzaki:2015che,Bian:2015kjt} because our scenario provides, at the same time, a composite pNGB Higgs and top partners (thus addressing the hierarchy problem), as well as minimality (only the WZW interactions are needed).

In this letter we explore the theory space of composite Higgs models with top partners based on an underlying fundamental dynamics, and find that all models predict a $\U(1)$ singlet pNGB potentially compatible with the recent di-photon hint.
We propose how to use this hint in combination with other di-boson searches in order to delineate the best candidate for composite higgs models with top-partners and demonstrate how it allows to severely reduce the theory space of the models under consideration and to make precise and testable prediction  for other signatures from surviving models. This strategy is also relevant in case the di-photon excess turns out to be a statistical fluctuation in the data, as other excesses may appear in the data.
 
\section{Model discussion}

The models we discuss in this letter are based on a confining $G_{HC}$ with two species of fermions, $\psi$ and $\chi$, which transform under two independent representations. We refer to $\psi$ as the colourless fermions which produce the Higgs as a pNGB, after condensation occurs. The number of such fermions $N_\psi$ (defined as the number of Weyl spinors) depends on their representation and the coset generated by their condensation: we can distinguish 3 classes of cosets, and for each we  consider the minimal case that contains a composite Higgs.
\begin{itemize}
\item[-] Real representation: the condensate breaks $\SU(N_\psi)\to \SO (N_\psi)$. The minimal case corresponds to $N_\psi = 5$, with 5 components transforming under the custodial $\SU(2)_L \times \SU(2)_R$ of the SM as a  $(2,2)$ plus a singlet.
\item[-] Pseudo-real representation: in this case, Dirac fermions need to be considered, and the coset arises as $\SU(N_\psi)/\Sp (N_\psi)$ with even $N_\psi$ and minimal case for $N_\psi = 4$. The 4 Weyl spinors transform as $(2,1) \oplus (1,2)$. 
\item[-] Complex representation: Dirac fermions are needed to avoid gauge anomalies ($N_\psi$ is even), and the coset is $\SU(N_\psi/2)^2/\SU(N_\psi/2)$. The minimal case corresponds to $N_\psi = 8$. The 4 Dirac fermions transform as $(2,1) \oplus (1,2)$. 
\end{itemize}
The top partners arise as fermionic bound states of the form $\psi \psi \chi$ or $\psi \chi \chi$, thus $\chi$ must carry both colour and an additional U(1) charge to fit the hypercharge of the SM quarks. Thus, the minimal $N_\chi$ is 6, comprising a Dirac triplet of colour. If the HC representation of $\chi$ is real (pseudo-real), than both colour and the additional charge $\U(1)_X$ are embedded in the coset $\SU(6)/\SO(6)$ ($\SU(6)/\Sp(6)$), while for complex representation colour is the unbroken group $\SU(3)^3/\SU(3)$ and the charge is the anomaly-free $\chi$-baryon number. Note that the assignment of the charge under $\U(1)_X$ is fixed by the HC representations of the two fermions.
A full list of the possible models~\cite{Ferretti:2013kya,Ferrettitalk} is shown in Table~\ref{table:classification}.

\begin{table}[t]
\begin{tabular}{|l|cc|c|c|c|}
\hline
$G_{HC}$ & $\psi$ & $\chi$ & EW  & Colour & $X$\\
\hline
Sp($2N_c$), $2 \leq N_c \leq 18$ & {\bf F} & {\bf A} & \multirow{2}{*}{$\frac{\SU(4)}{\Sp(4)}$} & \multirow{2}{*}{$\frac{SU(6)}{\SO(6)}$} & $2/3$\\
SO(11), SO(13) & {\bf Spin} & {\bf F} & & & $2/3$ \\
\hline
Sp($2N_c$), $N_c \geq 2$ & {\bf A} & {\bf F} & \multirow{3}{*}{$\frac{\SU(5)}{\SO(5)}$} & \multirow{3}{*}{$\frac{\SU(6)}{\Sp(6)}$} & $1/3$\\
Sp($2N_c$), $N_c \geq 6$ & {\bf Adj} & {\bf F} &  & & $1/3$ \\
SO(11), SO(13) & {\bf F} & {\bf Spin} & & & $1/3$\\
\hline
SO(7), SO(9) & {\bf Spin} & {\bf F} & \multirow{4}{*}{$\frac{SU(5)}{\SO(5)}$} & \multirow{4}{*}{$\frac{\SU(6)}{\SO(6)}$} & $2/3$\\
SO(7), SO(9) & {\bf F} & {\bf Spin} &  & & $1/3$\\
SO($N_c$), $N_c \geq 15$ & {\bf Adj} & {\bf F} & & & $1/3$ \\
SO($N_c$), $N_c \geq 55$ & {\bf S} & {\bf F} & & & $1/3$ \\
\hline
SU(4) & {\bf A} & {\bf F} & \multirow{2}{*}{$\frac{SU(5)}{\SO(5)}$} & \multirow{2}{*}{$\frac{{\SU(3)}^2}{\SU(3)}$} & $1/3$ \\
SO(10), SO(14) & {\bf F} & {\bf Spin} & &  & $1/3$ \\
\hline
SU(4) & {\bf F} & {\bf A} & \multirow{2}{*}{$\frac{{\SU(4)}^2}{\SU(4)}$} & \multirow{2}{*}{$\frac{\SU(6)}{\SO(6)}$} & $2/3$  \\
SO(10) & {\bf Spin} & {\bf F} &  &  & $2/3$ \\
\hline
SU(7) & {\bf F} & {\bf A$_3$} & \multirow{5}{*}{$\frac{{\SU(4)}^2}{\SU(4)}$} & \multirow{5}{*}{$\frac{{\SU(3)}^2}{\SU(3)}$} & $1/12$  \\
SU($N_c$), $N_c \geq 5$ & {\bf F} & {\bf A} & &  & $2/3$ \\
SU($N_c$), $N_c \geq 5$ & {\bf F} & {\bf S} & & & $2/3$ \\
SU($N_c$), $N_c \geq 5$ & {\bf A} & {\bf F} & &  & $1/12$\\
SU($N_c$), $N_c \geq 8$ & {\bf S} & {\bf F} & & & $1/12$ \\
 \hline
\end{tabular} \caption{The complete list of theories. The HC representations are: {\bf F} fundamental, {\bf S} 2-index symmetric, {\bf A} 2-index anti-symmetric, {\bf A$_3$} 3 index anti-symmetric, {\bf Adj} adjoint, {\bf Spin} spinorial of SO. The last column contains the $\U(1)_X$ charge assignment.}\label{table:classification}
\end{table}

\subsection{Couplings of the U(1) pNGBs}

Each model contains two $\U(1)$s that are spontaneously broken by the condensates: one associated to the $\psi$ fermions ($\U(1)_\psi$) and one to the $\chi$ fermions ($\U(1)_\chi$). In the cases with complex representation, the $\U(1)$ corresponds to the ``axial'' one, as the ``vector'' one is unbroken and anomaly-free.
However, one combination of the two has an anomaly with the $G_{HC}$, like the $\eta'$ in QCD, and will thus acquire a large mass of order $\Lambda_{HC}$ via instanton effects. The anomaly free $\U(1)$, which  is associated to the light pNGB, is defined by the following charge assignment to the two species of fermions:
\begin{equation}
q_\psi = N_\chi T_\chi\,, \quad q_\chi = - N_\psi T_\psi\,,
\end{equation}
where $T_{\psi,\chi}$ is the Dynkin index of the HC representation, and $N_{\psi,\chi}$ is the multiplicity of the fermions ($N_\chi = 6$, and $N_\psi = 4, 5, 8$ depending on the coset).

If we call $ a $ the pNGB of the spontaneously broken global U(1), its couplings to the gauge bosons via the WZW term can be parametrised as:
\begin{equation}
\mathcal{L} \supset \frac{g^2_i}{32 \pi^2} \frac{\kappa_i}{f_ a }\  a \ \epsilon^{\mu \nu \alpha \beta} G^i_{\mu \nu} G^i_{\alpha \beta}\,,
\end{equation}
where all the SM gauge groups are included (leading to 3 parameters: $\kappa_g$, $\kappa_W$ and $\kappa_B$).
Following the formalism used in~\cite{Cai:2015bss}, the coefficients can be computed by first writing the couplings of the two independent $\U(1)_{\psi,\chi}$. These coefficients only depend on the coset for the $\psi$ and $\chi$ condensates. We find
\begin{eqnarray}
\kappa_W = \kappa_B = d_\psi\,, & \mbox{for} & \frac{\SU(4)}{\Sp(4)}\,, \\
\kappa_W = \kappa_B = 2\ d_\psi \,, & \mbox{for} & \frac{\SU(5)}{\SO(5)}\; \mbox{and}\, \frac{\SU(4)^2}{\SU(4)}\,, \\
\begin{array}{l} \kappa_g = 2\ d_\chi\,,\\
\kappa_B = 12\ X^2\ d_\chi\,,  \end{array} & \mbox{for} & \mbox{all colour cosets}\,,
\end{eqnarray}
while coefficients not indicated above vanish. Here, $d_{\psi,\chi}$ are the dimensions of the HC representations.
The couplings of the pNGB can thus be written as
\begin{equation}
\frac{\kappa_i}{f_ a } = \frac{q_\psi \kappa_i^\psi + q_\chi \kappa_i^\chi}{\sqrt{q_\psi^2 f_\psi^2 + q_\chi^2 f_\chi^2}}\,,
\end{equation}
where $f_{\psi,\chi}$ are the two decay constants associated to the two $\U(1)_{\psi,\chi}$ breaking. Note that a single physical scale enters the couplings of $ a $: for future reference, we define
\begin{equation}
f_ a  = \sqrt{\frac{q_\psi^2 f_\psi^2 + q_\chi^2 f_\chi^2}{q_\psi^2 + q_\chi^2}}\,,
\end{equation}
as the physical scale associated to the couplings of $ a $, as its value is always between the two scales $f_{\psi,\chi}$.

With the above couplings, we can compute the branching ratios (BR) into pairs of gauge bosons, $gg$, $W^+ W^-$, $ZZ$, $Z\gamma$ and $\gamma \gamma$: the BRs are independent of the scale $f_ a $, and thus it only depend on the number of HC colours $N_c$ and the representation of the underlying fermions. In other words, they are a stark prediction of this class of models.
For convenience in comparing to experimental constraints, we define
\begin{equation}
R_{VV'} \equiv \frac{BR ( a  \to VV')}{BR( a  \to \gamma\gamma)}\,.
\end{equation}
We now assume that the $ a $ has a mass of 750 GeV, and is responsible for the di-photon excess reported at the LHC.
The only free parameter in the model, $f_ a $, can be determined by reproducing the cross section associated to the $\gamma \gamma$ signal. This value also determines the total width of the resonance. Using the connection to the signal rates, the total width can be written as:
\begin{equation}
\Gamma_{tot} = \sigma_{\gamma\gamma}^{\rm exp} \times \frac{\Gamma_{gg}}{\sigma (gg\to a )} \times \frac{(1+\sum_{VV'} R_{VV'})^2}{R_{gg}}\,,
\end{equation}
where $\sigma_{\gamma\gamma}^{\rm exp}$ is the experimental cross section, the second factor only depends on the mass of the resonance as the coupling dependence cancels out, and the last factor depends on the model.
This formula relies on the fact that the main production channel is gluon fusion, which is always true in the class of models considered here, for which, as we have checked, the contribution from sub-dominant vector-boson fusion is below per mille level.

\subsection{Implications of the  experimental constraints}

There are several experimental bounds relevant to test the consistency of the theory
with a di-photon  excess around 750 GeV.
First of all, using results of ATLAS~\cite{ATLAS} we  have estimated about 15  events above the background
in the excess region (from Fig.~1 of \cite{ATLAS}), which implies that the di-photon cross section at 13 TeV is 
$\sigma^{\gamma\gamma}_{13}\equiv \sigma_{13}\left(gg\rightarrow  a \rightarrow \gamma\gamma\right)\sim 
15/(3.2 \mbox{ fb}^{-1}\times 0.5) \simeq 10$~fb 
for the resonance production and decay. We used here 3.2 fb$^{-1}$
integrated luminosity and a (conservatively) estimated $50\%$ signal efficiency for a gluon-gluon fusion produced resonance -- the  production process under consideration.
 The di-photon search at CMS Run~I  \cite{Khachatryan:2015qba} imposes a bound of $\sigma^{\gamma\gamma}_{8} \lesssim 1.5\, (2.5)$~fb for a $\Gamma_a  < 0.1$~GeV ($\Gamma_a  < 75$~GeV) resonance. From Monte-Carlo simulations of the signal we find that  $ a $ production from gluon fusion is increased by a factor of 
\begin{equation}
\xi = \frac{\sigma_{13} (gg \to  a )}{\sigma_8 (gg\to  a )} \simeq 4.6 \label{eq:xi}
\end{equation}
for a 750 GeV resonance in Run~II as compared to Run~I, such that the Run~I di-photon search bounds are in mild tension with a $\sim 10$~fb narrow width di-photon resonance.

While experimental limits for other channels (di-jet and di-bosons)  are not available at Run II, yet, a 750 GeV resonance is constrained by the Run~I searches. According to the decay channels of the composite scalar under study, the relevant bounds on the production cross sections times branching ratio are the following: 
$\sigma^{gg}_{8} \lesssim 3$~pb \cite{Aad:2014aqa,CMS:2015neg}, $ \sigma^{WW}_{8} \lesssim 40$~fb \cite{Aad:2015agg} , $\sigma^{ZZ}_{8} \lesssim 12$~fb \cite{Aad:2015kna}, and  $\sigma^{Z\gamma}_{8} \lesssim 4$~fb \cite{Aad:2014fha}.\footnote{For the bounds we use the ATLAS or CMS search which yields the stronger constraint for the respective channel. Where available, we use bounds provided for gluon-fusion produced, scalar, narrow resonances, which resemble our di-photon resonance candidate most closely. The bounds are estimates as the corresponding ATLAS and CMS studies have been performed for BSM candidates. 
For a similar estimate of bounds {\it c.f.} {\it e.g.} \cite{Franceschini:2015kwy}.} Using the factor $\xi$ from Eq.~(\ref{eq:xi}) 
and assuming that the di-photon signal at Run~II originates from a 750~GeV resonance with 10 fb production cross section, the Run~I bounds can be translated into bounds on the 
ratios of $ a $ branching fractions,
\begin{equation}
R_{gg}\lesssim 1400, \ R_{WW}\lesssim 19, \ R_{ZZ}\lesssim 6, \ R_{Z\gamma} \lesssim 2\, ,
\label{eq:exp}
\end{equation}
where we used
\begin{equation}
R_{VV'}= \frac{Br( a \to VV')}{Br( a \to \gamma\gamma )}=\frac{\sigma_8 (gg \to  a  \to VV')}{\sigma_{13} (gg \to  a  \to \gamma\gamma)} \xi .
\end{equation}

The above values are based on the assumption of a 10 fb di-photon signal. A cross section of 5 fb is still within a 2 $\sigma$ statistical fluctuation, so that in the next section we call a model ``disfavoured as an explanation for  the 750  GeV di-photon resonance'' if the branching fractions exceed any of  the bounds of Eq.~(\ref{eq:exp}) by more than a factor of 2. We do not discuss bounds from the decay into pairs of SM fermions, both leptons and quarks, because these channels are closed in our models. 

\subsection{Results}

In table \ref{table: candidates} we show the models, from the list given in Table~\ref{table:classification}, that can explain the excess and are not excluded at 95~$\%$ of CL by present data, summarised in Eq.(\ref{eq:exp}). In the last two columns we report the total width and the value of $f_ a $ necessary to reproduce a signal strength of $10$ fb, in units of GeV. The values in bold exceed the bounds in Eq.(\ref{eq:exp}) but are allowed if we include a $50\%$ fluctuation in the observed di-photon excess.
It is noteworthy that in all cases a very small width for the signal is predicted and that the cross sections into other di-boson final states can be uniquely determined. 
These models robustly predict correlated additional di-boson and di-jet signals.
This allows us to create the respective strategy  to delineate  the properties of the underlying theory in case the 750 GeV di-photon signal is confirmed by the new data  from  Run II, or even in the case if this excess would go away while another,  di-boson or/and di-jet  signature(s) would appear.

\begin{table}[t]
\begin{tabular}{|lr|cccccc|}
\hline
& &$ R_{\text{WW}} $&$ R_{\text{ZZ}} $&$ R_{\text{Z$\gamma $}} $&$ R_{\text{gg}}
$&$ \Gamma _{\text{tot}} $&$ f_ a $\\\hline
  \text{SU(7)} &$({\bf F},{\bf A_3})$& $9.5$ &$ 3.0$ &$ 0.8 $&$ 140 $&$ 0.4 $& 2900 \\
 \text{SU(5)} &$({\bf A},{\bf F})$&$ 10 $&$ 3.2 $&$ 0.91 $&$ 1300$ &$ 3.2 $& 830 \\
 \hline
 \text{SO(11)} & $({\bf Spin},{\bf F})$& 4.4 & 0.51 & {\bf 3.5} & 500  & 0.8 & 2330 \\
  \text{SO(13)} &$({\bf Spin},{\bf F})$&$ 2.6 $&$ 0.2 $&$ {\bf2.6} $&$ 400 $ &$ 1.0 $& 4000 \\
\hline
 \text{SU(4)} & $({\bf A},{\bf F})$& {\bf 23} & {\bf 6.6} & {\bf 3.4} & 960 &  1.7 & 680 \\
  \text{SO(7)} &$({\bf F},{\bf Spin})$& $ {\bf 20} $&$ 5.7 $&$ {\bf2.7} $&$ 600$  &$ 1.5$ & 1300 \\
  \text{SO(9)} &$({\bf F},{\bf Spin})$& $ 16$ &$ 4.8$ &$ 2.0$ &$ 300$  &$ 0.8$ & 2200 \\
  \text{SO(10)} &$({\bf F},{\bf Spin})$& 15 & 4.6 & 1.8 & 227 & 0.6 & 2500 \\
  \text{SO(11)} &$({\bf F},{\bf Spin})$& 15 & 4.3 & 1.7& 180 & 0.4 & 2900 \\
   \text{SO(13)} &$({\bf F},{\bf Spin})$ & 13 & 4.1 & 1.5 & 120 & 0.3 & 3500 \\
   \text{SO(14)} &$({\bf F},{\bf Spin})$ & 13 & 4.0 & 1.4 & 99 & 0.2 & 3800 \\
\hline
\end{tabular} 
\caption{List of models that can explain the di-photon excess and are compatible with present data. The models are grouped according to the Higgs coset: $\SU(4)^2/\SU(4)$ for the top block, $\SU(4)/\Sp(4)$ for the second block, and $\SU(5)/\SO(5)$ for the bottom one. Values for $\Gamma _{\text{tot}}$ and $f_ a $ are given in GeV.}
 \label{table: candidates}
\end{table}

It is also interesting to notice that the scale $f_ a $ necessary to reproduce the di-photon cross section always falls in the TeV range, as one would naturally expect in models of composite Higgs.  For models based on $\SU(N_c)$ with ({\bf A}, {\bf F}) and ({\bf S}, {\bf F}) with large $N_c$, the values of the ratios are also compatible with the constraints, however a very large $f_ a $ is necessary to explain the di-photon rate. Therefore we do not consider this possibility any further.
The model studied in~\cite{Ferretti:2014qta}, based on $\SU(4)$ with ({\bf A}, {\bf F}) passes the bounds, even though there is a tension in the $WW$ rates, while the model in~\cite{Barnard:2013zea}, based on $\Sp(2N_c)$ with ({\bf F}, {\bf A}), cannot account for the signal.

\section{Conclusion}

In conclusion, we study  a complete set of models that work as underlying dynamics for models of composite Higgs with composite top partners. 
We point out that all models from this class contain a Standard Model singlet pNGB which couples to Standard Model gauge bosons through Wess-Zumino-Witten terms which is tested by  di-boson and di-jet searches at the LHC.
Within each model, the branching ratios of the pNGB into the various di-boson final states is fixed by the quantum
numbers of the constituent fermions. Assuming that the recently observed di-photon excess at 750 GeV results from the
production and decay of the pNGB, we show that the majority of models under consideration is ruled out by the LHC Run I
di-boson searches. The models which can explain the di-photon excess without being excluded by other di-boson searches are
shown in Table \ref{table: candidates}, together with their predictions for signal ratios in diboson channels, the pNGB
decay width, and the associated decay constant. If the di-photon signal is confirmed, future di-boson searches can further
discriminate between these candidate models in the near future. Even if the di-photon resonance is not confirmed, but a
different diboson resonance excess is found, the strategy we propose in  this letter can be applied to it in order to
discriminate between the models  since the presence of di-boson signals around the TeV scale is a robust prediction of this
class of models which deserve further explorations at the theoretical, phenomenological and experimental level.

\section*{Acknowledgements} 
We acknowledge support from the Franco-Korean {\it Partenariat Hubert Curien} (PHC) STAR 2015, project number 34299VE, and thank the France-Korea Particle Physics Lab (FKPPL) for partial support.
AB acknowledges partial support from the STFC grant ST/L000296/1,
the NExT Institute , Royal Society Leverhulme Trust Senior Research Fellowship LT140094 and
Soton-FAPESP grant.
GC and HC acknowledge partial support from the Labex-LIO (Lyon Institute of Origins) under grant ANR-10-LABX-66 and FRAMA (FR3127, F\'ed\'eration de Recherche ``Andr\'e Marie Amp\`ere'').
TF and HS were supported by the Basic Science Research Program through the National Research Foundation of Korea (NRF) funded by the ministry of Education, Science and Technology (No. 2013R1A1A1062597).


\end{document}